\documentclass[useAMS,usegraphicx]{mn2e}

\newcommand{\Msun}{\ensuremath{\,{\rm M}_\odot}}            
\newcommand{\Rsun}{\ensuremath{\,{\rm R}_\odot}}            
\newcommand{\Teff}{\ensuremath{T_{\rm eff}}}                
\newcommand{\logg}{\ensuremath{\log g}}                     
\newcommand{\Vsys}{\ensuremath{V_\gamma}}                   
\newcommand{\Veq}{\ensuremath{V_{\rm eq}}}                  
\newcommand{\Vsync}{\ensuremath{V_{\rm synch}}}             
\newcommand{\EBV}{\ensuremath{E_{B-V}}}                     
\newcommand{\Eby}{\ensuremath{E_{b-y}}}                     
\newcommand{\kms}{\,km\,s$^{-1}$}                           
\newcommand{\cms}{\,cm\,s$^{-2}$}                           
\newcommand{\ion}[2]{{#1}\,{\sc {\small{#2}}}}              
\newcommand{\etal}{\mbox{et\,al.}}
\newcommand{\mc}[1]{\multicolumn{2}{c}{#1}}

\title[Eclipsing binaries in open clusters. II. V453\,Cyg in NGC\,6871]
      {Eclipsing binaries in open clusters. \\ II. V453\,Cyg in NGC\,6871
       \thanks{Based on observations made with the Isaac Newton Telescope operated on the island of La Palma 
               by the Isaac Newton Group in the Spanish Observatorio del Roque de los Muchachos of the 
               Instituto de Astrofis\'\i ca de Canarias}}

\author[J.\ Southworth, P.\ F.\ L.\ Maxted and B.\ Smalley]
       {J.\ Southworth$^1$\thanks{Email addresses: jkt,pflm,bs@astro.keele.ac.uk}, 
        P.\ F.\ L.\ Maxted$^1$\footnotemark[2] and B.\ Smalley$^1$\footnotemark[2] \\
        $^1$Department of Physics and Chemistry, Keele University, Staffordshire, ST5 5BG, UK}

\begin{document} \maketitle 

\begin{abstract}
We derive absolute dimensions of the early B-type detached eclipsing binary V453\,Cygni (B0.4\,IV + B0.7\,IV, $P=3.89$\,d), a member of the open cluster NGC\,6871. From the analysis of new, high-resolution, spectroscopy and the $UBV$ light curves of Cohen (1974) we find the masses to be $14.36 \pm 0.20$\Msun\ and $11.11 \pm 0.13$\Msun, the radii to be $8.55 \pm 0.06$\Rsun\ and $5.49 \pm 0.06$\Rsun, and the effective temperatures to be $26\,600 \pm 500$\,K and $25\,500 \pm 800$\,K for the primary and secondary stars, respectively. The surface gravity values of $\logg = 3.731 \pm 0.012$ and $4.005 \pm 0.015$ indicate that V453\,Cyg is reaching the end of its main sequence lifetime. We have determined the apsidal motion period of the system to be $66.4 \pm 1.8$\,yr using the technique of Lacy (1992) extended to include spectroscopic data as well as times of minimum light, giving a density concentration coefficient of $\log k_2 = -2.226 \pm 0.024$. Contaminating (third) light has been detected for the first time in the light curve of V453\,Cyg; previous analyses without this effect systematically underestimate the ratio of the radii of the two stars. The absolute dimensions of the system have been compared to the stellar evolution models of the Granada, Geneva, Padova and Cambridge groups. All model sets fit the data on V453\,Cyg for solar helium and metal abundances and an age of $10.0 \pm 0.2$\,Myr. The Granada models also agree fully with the observed $\log k_2$ once general relativistic effects have been accounted for. The Cambridge models with convective core overshooting fit V453\,Cyg better than those without. Given this success of the theoretical predictions, we briefly discuss which eclipsing binaries should be studied in order to further challenge the models.
\end{abstract}



\section{Introduction}        \label{introduction}   

Theoretical models of high-mass stars are very difficult to construct, due to a number of poorly understood astrophysical phenomena which become important at higher stellar masses. Convective core overshooting is an essential ingredient of stellar models and can have a large effect on the lifetimes and luminosities of high-mass stars. The degree of overshooting may depend on metallicity (Cordier \etal\ 2002, Palmieri \etal\ 2002) and stellar mass (Young \etal\ 2001). Stellar rotation is known to cause enhanced mixing in the interiors of stars which can mimic a small amount of convective overshooting and strongly affect the evolution of stars (Maeder \& Meynet 2000). Mass loss is also an important effect (Maeder 1997) but is usually represented in theoretical models by a parameterisation such as that of Reimers (1975) (Woo \etal\ 2003).

Until recently there was a discrepancy between masses derived from spectroscopic observations and inferred from evolutionary models (Herrero \etal\ 1992). Recent advances in theoretical modelling, and the increased sophistication of observational techniques, appears to have removed this mass discrepancy (Hilditch, 2004), but confirmation of this requires detailed comparisons between models and as many physical properties of individual stars as possible.


Detached eclipsing binaries (dEBs) are a vital source of observational data on the radii and masses of high-mass stars (Andersen 1991) but studies of such stars are hindered by the rarity of useful systems, the difficulty of measuring accurate velocities from spectra of rapidly-rotating multiple stars (Sana, Rauw \& Gosset 2001), and the large amount of telescope time needed to obtain light curves which are complete over the whole orbital period. This has caused a shortage of data on eclipsing binaries with masses above 10\Msun\ and sufficiently large orbital periods to be detached and therefore to have evolved as single stars.

Whilst dEBs can provide accurate observed stellar masses, radii, effective temperatures and equatorial rotational velocities, the ages, chemical compositions and distances of such systems are not, in general, directly observable. This means that when fitting observed quantities to theoretical predictions, there are several free parameters which can be adjusted to find the best fit to the observational data. This makes it very difficult to assess the success of important but more subtle physical effects such as rotational mixing, mass loss, convective overshooting and even the type of turbulent convective theory. 

Eclipsing binaries in open clusters and associations can be analysed to derive accurate masses, radii, temperatures and rotational velocities. Their membership of a star cluster can also provide knowledge of the age, chemical composition, and distance of the dEB, allowing a more complete physical description of a single stellar system. This reduces the number of free parameters adjustable when fitting theoretical models to observations, which allows a discriminate test of the representation of different physical effects and indeed the success of different sets of stellar evolutionary models. For example, from the analysis of two main sequence dEBs which are members of the open cluster h\,Persei (NGC\,869), Southworth, Maxted \& Smalley (2004, hereafter Paper\,I) were able to provide a first estimate of the bulk metallicity of one of the most important and well-studied open clusters in the Northern Hemisphere.

\subsection{V453\,Cygni in NGC\,6871}     \label{v453cyg}

\begin{table} \begin{center} 
\caption{\label{tablephotdata} Identifications, location, and combined photometric parameters for the V453\,Cygni eclipsing system. \newline {\bf References:} (1) Cannon \& Pickering (1923); (2) Argelander (1903); (3) H{\o}g \etal\ (1998); (4) Hoag \etal (1961); (5) Popper (1980); (6) Zakirov (1992); (7) Cohen (1969); (8) Reimann (1989).} 
\begin{tabular}{lr@{}lr} \hline \hline
                        &     & V453 Cygni            & Reference \\ \hline
Henry Draper number     &     & HD 227696             & 1   \\
Bonner Durchmusterung   &     & BD\,+35\degr 3964     & 2   \\ 
Hoag number             &     & NGC 6871 13           & 3   \\ \hline 
$\alpha_{2000}$         &     & 20 06 34.967          & 4   \\
$\delta_{2000}$         &   + & 35 44 26.28           & 4   \\
Spectral type           &   & B\,0.4\,IV + B\,0.7\,IV & 5   \\ \hline
$V$                     &     & 8.285                 & 6   \\
$B-V$                   &   + & 0.179                 & 6   \\
$U-B$                   & $-$ & 0.61                  & 6   \\ 
$V-R$                   &   + & 0.254                 & 6   \\
$\beta$                 &     & 2.590                 & 7,8 \\
\hline \hline \end{tabular} \end{center} \end{table}

\begin{table*} \begin{center} 
\caption{\label{pubspecorbits} Published spectroscopic orbits of V453\,Cygni. BMM97 originally fitted their data with a circular orbit. We have refitted their radial velocities with an eccentric orbit to increase the accuracy of our determination of the apsidal motion (see section~\ref{perioddet} for details). A colon after a number indicates that it is uncertain. Quantities without quoted errors or a colon were not determined by that investigation. When quantities are given separately for each star we have quoted a weighted mean of the two values. Symbols have their usual meanings. Times are written as (HJD $-$ 2\,400\,000).
\newline $^*$\,The reference time, $T_0$, refers to a time of periastron passage, not a time of minimum light.} 
\begin{tabular}{l r@{\,$\pm$\,}l r@{\,$\pm$\,}l r@{\,$\pm$\,}l r@{\,$\pm$\,}l r@{\,$\pm$\,}l r@{\,$\pm$\,}l } \hline \hline
& \mc{Pearce}  & \mc{Abt, Levy and} & \mc{Popper and} &\mc{Simon and}   & \mc{BMM97}      & \mc{BMM97}          \\
& \mc{(1941)}  & \mc{Gandet (1972)} & \mc{Hill (1991)}&\mc{Sturm (1994)}& \mc{}           & \mc{(our solution)} \\ \hline 
$P$ (days)     
& \mc{3.87972} & \mc{3.8890}        & \mc{3.8898128}  & \mc{3.88982309} &\mc{3.8898128}   &\mc{3.889825}        \\
$T_0$ (HJD)    
&30231.0843&0.0543$^*$ & \mc{40495.027$^*$} & \mc{39340.099} & \mc{36811.7296} &48141.82&0.01$^*$& 48500.64 & 0.66$^*$ \\
$K_{\rm A}$ (\kms) 
& 181.8 & 1.13 & \mc{152:}          & 171 & 1.5       & 171.7 & 2.9     & 173.2 & 1.3     & 173.7 & 1.4         \\
$K_{\rm B}$ (\kms) 
& 237.4 & 2.78 & \mc{}              & 222 & 2.5       & 223.1 & 2.9     & 213.6 & 3.0     & 212.4 & 3.4         \\
$e$ 
& 0.07 & 0.007 & \mc{0.05:}         &\mc{0.0}         &\mc{0.0}         & \mc{0.0}        & 0.011 & 0.015       \\
$\omega$ (degrees)
& 175.2 & 5.06 & \mc{99:}           & \mc{}           & \mc{}           & \mc{}           & 88.6 & 6.0          \\
\Vsys\ (\kms)     
& $-$15.0&0.94 & \mc{$-$22.7:}      &\mc{$-$14}       & \mc{$-$7:}      & $-17.6$ & 1.0   & $-$18.0 & 1.6       \\
\hline \hline \end{tabular} \end{center} \end{table*}

V453\,Cygni is a high-mass dEB with an orbital period of 3.89\,days. Its membership of the young open cluster NGC\,6871 means that its age and distance can be found independently. The primary component of V453\,Cyg is approaching the terminal age main sequence (TAMS) and its large radius causes the eclipses to be total, allowing a very accurate determination of the radii of both stars. 
Table~\ref{tablephotdata} contains identifications and some photometric properties of the system.

The eclipsing nature of V453\,Cyg was discovered by Wachmann (1939) and an early spectroscopic orbit was calculated by Pearce (1941). A period study by Cohen (1971) provided a determination of the orbital longitude of periastron, $\omega$, inconsistent with that derived by Pearce. In a period study by Wachmann (1973) this was correctly interpreted as apsidal motion. Wachmann derived an apsidal period of $U = 72$ years using measurements of the time differences between several groups of adjacent primary and secondary eclipses. A more recent period study, including parabolic and periodic terms, was undertaken by Rafert (1982). 

Excellent photoelectric $UBV$ light curves were observed by Wachmann (1974) and analysed using the Russell-Merrill method (Russell \& Merrill 1952) which involves the procedure of rectification. His work contains a plot of the light curves adjusted for the effects of orbital eccentricity and apsidal motion using parameters updated from that of his previous work, but the data themselves have so far been unobtainable. It is possible that they are in an unlabelled file in the IAU Variable Star Archives (Breger 1988), but no record of them exists at Hamburg Observatory, where the light curves were observed (A.\ Reiners, private communication).

Cohen (1974) published complete photoelectric $UBV$ light curves which contain fewer datapoints and more observational scatter than those of Wachmann (1974). He analysed these using the Russell-Merrill method but stated that his observations were not definitive. They have since been analysed by Cester \etal\ (1978) using the light curve analysis code {\sc wink} (Wood 1971). This is the only previous photometric study to use modern techniques.

A recent investigation using photoelectric $UBVRI$ light curves has been published by Zakirov (1992). He analysed his light curves using the ``direct machine method of Lavrov (1993)'', which is based on rectification. The results of the four photometric analyses of V453\,Cyg are substantially in agreement about the basic photometric parameters of the system.

Recent spectroscopic orbits have been published by Popper \& Hill (1991), Simon \& Sturm (1994) and Burkholder, Massey \& Morrell (1997, hereafter BMM97). These results are collected in Table~\ref{pubspecorbits}. Simon \& Sturm used seven spectra to demonstrate their spectral disentangling algorithm, which decomposes observed composite spectra into the separate spectra of two stars. The total secondary eclipse of V453\,Cyg allowed them to directly compare their disentangled primary spectrum with a spectrum observed during secondary eclipse. Disentangling can be used to determine accurate orbital semiamplitudes (Hynes \& Maxted 1998, Harries \etal\ 2003) but it is not clear if there is a robust method by which to estimate uncertainties in the derived quantities. BMM97 derived a good spectroscopic orbit from 25 high signal-to-noise spectra and compared the dEB to models to investigate the discrepancy at higher masses between models and observations. The rotational velocities of the components of V453\,Cyg were determined by Olson (1984) to be $107 \pm 9$\kms\ and $97 \pm 20$\kms\ for the primary and secondary stars respectively. A preliminary single-lined spectroscopic orbit was also given by Abt, Levy \& Gandet (1972).

An abundance analysis of V453\,Cyg was undertaken by Daflon \etal\ (2001) using both LTE and non-LTE calculations. The results suggest that V453\,Cyg has a slightly sub-solar metallicity. These authors derived an effective temperature of 29\,100\,K using the $Q$ parameter based on $UBV$ magnitudes (Johnson 1957), and a surface gravity of $\logg = 4.45$ (\cms) from profile fitting of the H$\gamma$ 4340\,\AA\ spectral line. Both values are larger than expected and inconsistent with previous analyses. Their surface gravity value is in fact somewhat larger than appropriate for the ZAMS, and is inconsistent with significant main sequence evolution.


\subsection{NGC\,6871}              \label{ngc6871}

The open cluster NGC\,6871 is a concentration of bright OB stars which forms the nucleus of the Cyg\,OB3 association (Garmany \& Stencel 1992). This makes it an important object for the study of the evolution of high-mass stars. The cluster itself has been studied photometrically several times but its sparse nature means determination of its physical parameters is difficult. 

$UBV$ photometry of the 30 brightest stars was published by Hoag \etal\ (1961). Crawford, Barnes \& Warren (1974) observed 11 stars in the Str\"omgren $uvby$ system and 24 stars in the Crawford $\beta$ system, finding significantly variable reddening and a distance modulus of 11.5\,mag. This $uvby\beta$ photometry was extended to 40 stars by Reimann (1989), who found reddening $\Eby$ with a mean value of 0.348\,mag and an intracluster variation of about 0.1\,mag. His derived distance modulus of $11.94 \pm 0.08$\,mag and age of 12\,Myr are both greater than previous literature values.

Massey, Johnson \& DeGioia-Eastwood (1995) conducted extensive $UBV$ CCD photometry of 1955 stars in the area of Cyg\,OB3. Their values of distance modulus, $11.65 \pm 0.07$, and of reddening, $\EBV = 0.46 \pm 0.03$\,mag with individual values between 0.04 and 1.11 mag, agree well with previous determinations. They find an age of 2 to 5 Myr for stars with spectral types earlier than B0 but give evidence for a significant spread of stellar ages in the cluster. Whilst the highest-mass unevolved cluster members have main sequence lifetimes of 4 to 5 Myr, NGC\,6871 contains evolved 15\Msun\ stars despite their main sequence lifetimes being of the order of 11\,Myr.


\section{Observations}  \label{observations}  

Spectroscopic observations were carried out in 2002 October on the 2.5\,m Isaac Newton Telescope (INT) on La Palma. The 500\,mm camera of the Intermediate Dispersion Spectrograph (IDS) was used with a holographic 2400\,{\it l}\,mm$^{-1}$ grating and EEV 4k\,$\times$\,2k CCD. From measurements of the full width half maximum (FWHM) of arc lines taken for wavelength calibration we estimate that the resolution is approximately 0.2\,\AA. The spectral windows chosen for observation were 4450--4715\,\AA\ (31 spectra) and 4230--4500\,\AA\ (12 spectra). Additional spectra were observed around H$\beta$ (4861\,\AA) to provide an additional temperature indicator for spectral analysis. The signal-to-noise ratio per pixel of the observed spectra is between 100 and 450.

Data reduction was undertaken using optimal extraction as implemented in the software tools {\sc pamela} and {\sc molly}
\footnote{{\sc pamela} and {\sc molly} were written by Dr.\ Tom Marsh and can be found at \texttt{http://www.warwick.ac.uk/staff/T.R.Marsh}} (Marsh 1989).


\section{Period determination and apsidal motion} \label{perioddet}      

\begin{table} \begin{center} 
\caption{\label{mintable} Times of minimum light of V453\,Cyg taken from the literature. The $O-C$ values refer to the difference between the observed and calculated values. \newline$^*$\,Rejected from the fit due to a large $O-C$ value.
\newline {\bf References:} (1) Wachmann (1973) photographic, (2) Wachmann (1973) photoelectric, (3) Cohen (1971) photoelectric, (4) R.\ Diethelm (see text) photoelectric, (5) B\'{\i}r\'o \etal\ (1998) CCD.}
\begin{tabular}{rlrrr} \hline \hline
Cycle       & Minimum time      & Adopted       & $O-C$       & Ref.      \\
number      &(HJD$-$2\,400\,000)& error         &             &           \\ \hline
$-$2790.0   & 28487.531         &    0.01       &    0.0026   & 1         \\
$-$2789.5   & 28489.435         &    0.01       &    0.0008   & 1         \\
$-$2608.0   & 29195.476         &    0.01       & $-$0.0028   & 1         \\
$-$2607.5   & 29197.371         &    0.01       & $-$0.0018   & 1         \\
$-$1501.0   & 33501.508         &    0.01       & $-$0.0020   & 1         \\
$-$1500.5   & 33503.414         &    0.01       & $-$0.0008   & 1         \\
$-$1482.0   & 33575.411         &    0.01       & $-$0.0054   & 1         \\
$-$1481.5   & 33577.328         &    0.01       &    0.0061   & 1         \\
$-$1390.0   & 33933.270         &    0.01       & $-$0.0084   & 1         \\
$-$1389.5   & 33935.195         &    0.01       &    0.0074   & 1         \\
  $-$65.0   & 39087.266         &    0.005      &    0.0039   & 2         \\
  $-$64.5   & 39089.242         &    0.005      &    0.0028   & 2         \\
   $-$7.5   & 39310.9552        &    0.005      & $-$0.0052   & 2         \\
   $-$7.0   & 39312.8702        &    0.005      & $-$0.0010   & 3         \\
   $-$5.5   & 39318.7347        &    0.005      & $-$0.0054   & 3         \\
   $-$5.0   & 39320.6497        &    0.005      & $-$0.0011   & 3         \\
     12.0   & 39386.7764        &    0.005      & $-$0.0010   & 3         \\
    178.0   & 40032.492         &    0.005      &    0.0068   & 2         \\
    178.5   & 40034.470         &    0.005      & $-$0.0015   & 2         \\
   1684.0   & 45890.5660        &    0.005      &    0.0016   & 4         \\
   2801.0   & 50235.4843$^*$    &    0.005      & $-$0.0426   & 5         \\
\hline \hline \end{tabular} \end{center} \end{table}

\begin{table} \begin{center} 
\caption{\label{specmintable} Spectroscopic data used in the apsidal motion analysis.
\newline {\bf References:} (1) Pearce (1941), (2) BMM97 (our solution).}
\begin{tabular}{r r@{\,$\pm$\,}l c r@{\,$\pm$\,}l c r} \hline \hline
Cycle       & \mc{Eccentricity}     & $O\!-\!C$     & \mc{$\omega$}     & $O\!-\!C$ & Ref.      \\
number      & \mc{}                 & ($e$)         & \mc{(degrees)}    & ($\omega$)&           \\ \hline
$-$2342.0   & 0.070 & 0.007         & 0.048         & 175.2 & 5.1       & 1.0       & 1         \\
2355.0      & 0.011 & 0.015         & 0.011         &  88.6 & 6.0       & 2.6       & 2         \\
\hline \hline \end{tabular} \end{center} \end{table}

\begin{table*} \begin{center} 
\caption{\label{apsmottable} Apdisal motion parameters for V453\,Cyg. To illustrate the final result we have also included the best-fitting parameters using only photometric data, and the results of Wachmann (1974). Times are written as (HJD $-$ 2\,400\,000).} 
\begin{tabular}{l ccc r@{\,$\pm$\,}l r@{\,$\pm$\,}l} \hline \hline
                                                      &\hspace{10pt}& Wachmann &\hspace{10pt}& \mc{This paper} & \mc{This paper}\\
                                                      && (1974) && \mc{(photometric only)} & \mc{(final results)} \\ \hline
Anomalistic period $P$ (days)                         && && 3.890426 & 0.000073 & 3.890450 & 0.000017 \\
Reference minimum time $T_0$                          && 36811.7296 && 39340.1011 & 0.0022 & 39340.0998 & 0.0019 \\
Orbital inclination $i$ (degrees)                     && && \mc{88.0 (fixed)} & \mc{88.0 (fixed)} \\
Orbital eccentricity $e$                              && 0.02 && 0.019 & 0.002 & 0.022 & 0.002 \\
Periastron longitude at $T_0$, $\omega_0$ (degrees)   && 309.2 && 313.2 & 6.7 & 309.7 & 3.1 \\
Apsidal motion rate $\dot\omega$ (degrees $P_s^{-1}$) && 0.0539 && 0.0556 & 0.0068 & 0.0579 & 0.0016 \\
Sidereal period $P_s$ (days)                          && 3.88982309 && 3.889824 & 0.000082 & 3.889825 & 0.000018 \\
Apsidal motion period $U$ (years)                     && 71 && 68.9 & 8.5 & 66.4 & 1.8 \\
\hline \hline \end{tabular} \end{center} \end{table*}

\begin{figure*} \includegraphics[width=\textwidth,angle=0]{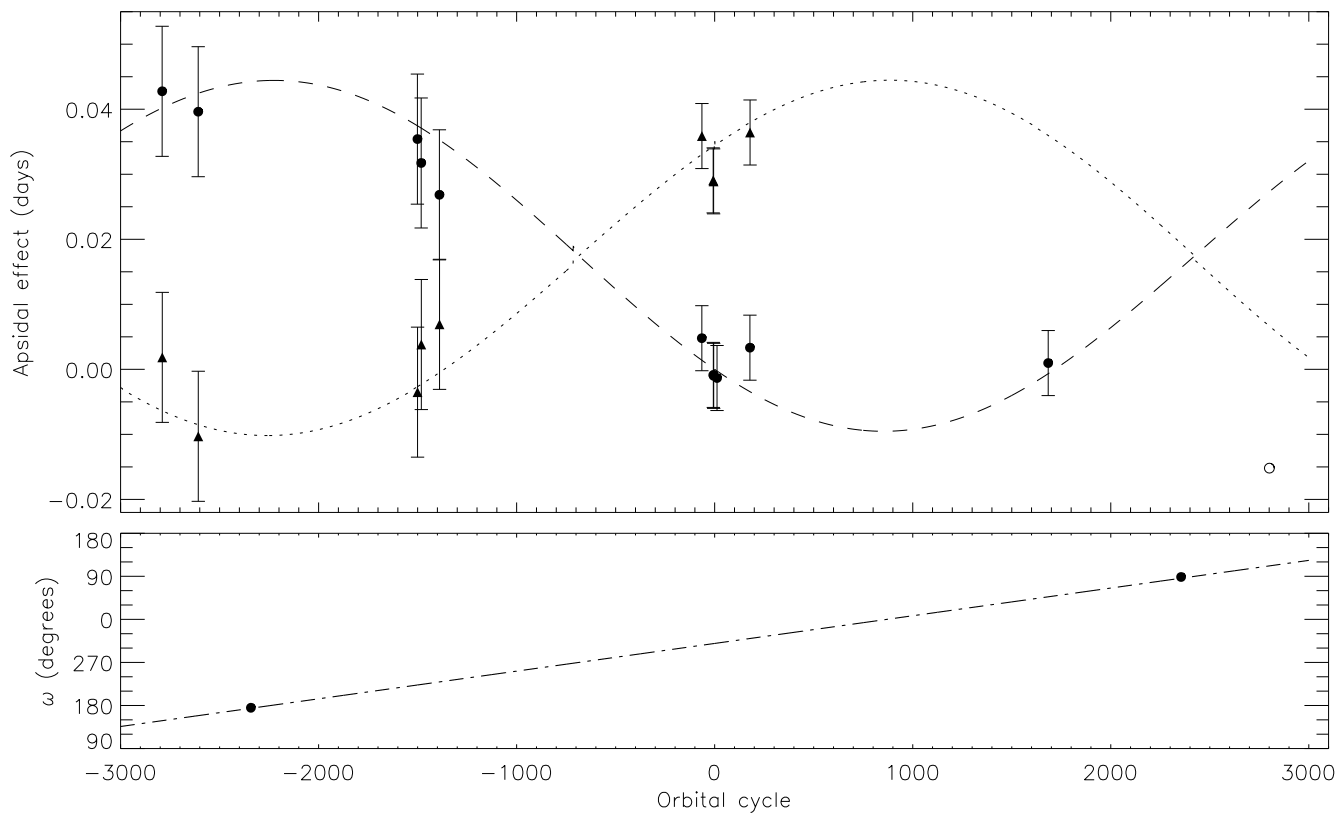}
\caption{\label{apsmotfigure} Representation of the best-fitting apsidal motion parameters. The upper panel shows the observed times of primary (circles) and secondary (triangles) minima, minus the expected times given by a linear ephemeris, compared to the best-fitting curves of primary (dashed) and secondary (dotted) minima. The open circle represents the rejected time of minimum of B\'{\i}r\'o \etal\ (1998). The lower panel shows the spectroscopically determined longitudes of periastron, $\omega$, and the change of $\omega$ over orbital cycle. Errors have only been shown if they are larger than the corresponding symbol.} \end{figure*}

A period study by Wachmann (1973, 1974) indicated fast apsidal motion with a period of $U = 71$\,yr from eighteen times of minimum light covering almost three thousand orbital cycles. Photographic minima exist dating back to the year 1902 but they are not of sufficient quality to improve the apsidal motion analysis (Ashbrook, unpublished, but tabulated in Cohen 1971). Zakirov (1991) states that his observations are not consistent with Wachmann's apsidal motion period.

Times of minima for inclusion in the apsidal motion analysis were taken from Cohen (1971), Wachmann (1973), R.\ Diethelm\footnote{Eclipsing Binaries Minima Database at \\ \texttt{http://www.oa.uj.edu.pl/ktt/index.html}}, and B\'{\i}r\'o \etal\ (1998). The method of Lacy (1992) was adopted to solve the apsidal motion equations without the use of approximations. Here the Levenberg-Marquardt non-linear least-squares fitting routine (Press \etal\ 1992) is used to relate the derived apsidal motion parameters to the observed times of minima. The authors' implementation of this method ({\sc apsmot}) uses subroutines very generously supplied by D.\ Holmgren (see Holmgren \& Wolf 1996).

It was immediately clear that the more recent times of minima were not in full agreement with each other or with the times of minima used by Wachmann (1973). In such cases, independent information is needed to decide which published observations are reliable and which are discrepant. For this reason we added to our {\sc apsmot} code the ability to include spectroscopic determinations of eccentricity, $e$, and longitude of periastron, $\omega$, in the overall fit. 

Values of $e$ and $\omega$ were taken from the spectroscopic studies by Pearce (1941) and BMM97. The radial velocity observations of BMM97 were originally fitted with a circular spectroscopic orbit so we have reanalysed the velocities (Table~\ref{pubspecorbits}) to determine $e$ and $\omega$, and assigned a cycle number corresponding to the approximate midpoint of the observations. We have fitted an eccentric orbit to the data using {\sc sbop}\footnote{Spectroscopic Binary Orbit Program written by Dr.\ Paul B.\ Etzel (\texttt{http://mintaka.sdsu.edu/faculty/etzel/}).} in single-lined Lehman-Filh\'es mode. There is a large correlation between $\omega$ and the ephemeris reference time, $T_0$, which makes both values somewhat uncertain. The solution of the secondary velocities did not converge without fixing the value of $T_0$ to that of the primary star's spectroscopic orbit.

The spectroscopic data of Pearce (1941) and BMM97 allowed us to identify the time of minimum of B\'{\i}r\'o \etal\ (1998) as being in disagreement with the other photometric data. This datapoint has been rejected from the apsidal motion solution, and the other data were assigned appropriate uncertainties. These data are given in Tables \ref{mintable} and \ref{specmintable} along with the assigned uncertainties and the observed minus calculated ($O-C$) values. The final solution is plotted against the data in Fig.~\ref{apsmotfigure} and given in Table~\ref{apsmottable}, where it is compared to the solution of Wachmann (1974) and to a solution without the inclusion of spectroscopic data. 

The time of minimum of B\'{\i}r\'o \etal\ (1998) has been confirmed by reanalysis and another unpublished time of minimum (I.\ B\'{\i}r\'o, private communication). If they are correct, they may indicate the existence on another effect on the times of minima, for example the light-time effect. Further data are needed to investigate this possibility.


\section{Spectral synthesis}   \label{specsynth}

The observed spectra were fitted to synthetic spectra, by the method of least squares, to derive the effective temperatures of the components of V453\,Cyg. Synthetic spectra were calculated using {\sc uclsyn} (Smith 1992, Smalley \etal\ 2001), Kurucz (1993) {\sc atlas9} model atmospheres and absorption lines from the Kurucz \& Bell (1995) linelist. The profiles for the 4387.93\,\AA\ and 4471.50\,\AA\ \ion{He}{i} lines were calculated using profiles from the work of Barnard \etal\ (1969) and Shamey (1969), with $\log gf$ values from the critical compilation of Wiese \etal\ (1966). The spectra were rotationally broadened as necessary and an instrumental broadening of FWHM = 0.2\,\AA\ was applied to match the resolution of the observations.

The primary star was analysed using a spectrum obtained during a total secondary eclipse. For the secondary star we used the spectra at quadrature to measure the lines and corrected them for dilution effects. The equivalent widths of the helium lines are given in Table~\ref{HeEW} and were used to obtain values of effective temperature (\Teff) by ensuring ionisation balance between \ion{He}{i} and \ion{He}{ii}, for assumed values of surface gravity and microturbulence velocity ($\xi_t$).

\begin{table} \begin{center} \caption{\label{HeEW} Equivalent widths of helium lines in the spectra of V453\,Cyg. These are true equivalent widths per individual star, after corrections for dilution due to the spectra being composite.}
\begin{tabular}{llccc} \hline \hline
Species           & Wavelength      &\hspace{10pt}& \multicolumn{2}{c}{Equivalent widths (\AA)}   \\
                  & of line (\AA)   &             & primary star          & secondary star        \\ \hline
\ion{He}{i}       & 4387.93         &             & 0.661                 & 0.890                 \\
\ion{He}{i}       & 4437.55         &             & 0.089                 & $\cdots$              \\
\ion{He}{i}       & 4471.50         &             & 0.992                 & 1.18                  \\
\ion{He}{ii}      & 4685.70         &             & 0.139                 & 0.057                 \\ 
\hline \hline \end{tabular}\end{center} \end{table}

Using the surface gravities found from the spectroscopic and photometric analyses in this work, we find the effective temperatures $\Teff = 26\,600 \pm 500$\,K for the primary star and $\Teff = 25\,500 \pm 800$\,K for the secondary star. These parameters imply that the atmospheres of these stars are helium-rich by about 0.25\,dex compared to solar. Further support for these \Teff\ and \logg\ values is given by the H$\gamma$ 4340\,\AA\ profiles; H$\gamma$ profiles with higher \logg\ (as found by Daflon \etal\ 2001) are too broad to fit the observations.

Using the $uvby\beta$ photometry from Hauck \& Mermilliod (1998) and the {\sc uvbybeta} and {\sc tefflogg} codes of Moon (1985), we have obtained de-reddened photometry and fits to the grids of Moon \& Dworetsky (1984). Values of $\Teff = 26\,710 \pm 800$\,K and $\logg = 3.78 \pm 0.07$ were obtained, in excellent agreement with the parameters of the primary star, which dominates the system light, obtained in this work.

Daflon \etal\ (2001) adopted $\Teff = 29\,100$\,K, $\logg = 4.45$ (\cms) and $\xi_t = 12$\kms\ in their detailed analysis of V453\,Cyg. The above ionisation balance analysis gives a $\Teff = 29\,200$\,K for their \logg\ and $\xi_t$. While this is in agreement with the \Teff\ they adopted, their value of \logg\ is not supported by our absolute stellar dimensions, our H$\gamma$ profile fitting and the $uvby\beta$ photometry, so we prefer cooler effective temperatures for the components of V453\,Cyg.


\section{Spectroscopic orbits} \label{todcor}   

\begin{table} \begin{center} \caption{\label{rvtable}
Radial velocities and $O-C$ values (in \kms) for V453\,Cyg calculated using {\sc todcor}. Weights are given in column ``Wt'' and were derived from the amount of light collected in that observation and were used in the {\sc sbop} analysis.}
\begin{tabular}{lrrrrr} \hline \hline
HJD $-$     & Primary  & $O-C$ & Secondary  & $O-C$ & Wt \\ 
2\,400\,000 & velocity &       & velocity   &       &    \\ \hline
52562.5002 & $-$188.9 &  $-$0.4 &    201.3 &  $-$9.3 & 1.0 \\
52564.3199 &    142.9 &  $-$4.0 & $-$218.9 &     4.2 & 1.0 \\
52566.2961 & $-$181.4 &     1.8 &    200.7 &  $-$3.0 & 0.7 \\
52566.3112 & $-$191.0 &  $-$6.7 &    204.5 &  $-$0.7 & 0.7 \\
52566.4356 & $-$191.5 &  $-$1.9 &    190.6 & $-$21.4 & 1.2 \\
52568.3471 &    150.8 &  $-$5.2 & $-$227.8 &     7.0 & 1.0 \\
52560.3187 &    137.6 &     3.5 & $-$206.5 &     0.1 & 1.0 \\
52560.3210 &    136.0 &     1.5 & $-$211.7 &  $-$4.6 & 1.0 \\
52560.3233 &    133.0 &  $-$1.8 & $-$217.9 & $-$10.5 & 1.0 \\
52560.3257 &    133.8 &  $-$1.3 & $-$218.2 & $-$10.3 & 1.0 \\
52560.3280 &    131.1 &  $-$4.3 & $-$211.7 &  $-$3.4 & 1.1 \\
52560.4091 &    142.5 &  $-$2.4 & $-$219.0 &     1.4 & 0.8 \\
52560.4114 &    146.4 &     1.2 & $-$221.5 &  $-$0.7 & 0.8 \\
52560.4137 &    149.3 &     4.0 & $-$225.9 &  $-$4.8 & 0.8 \\
52560.4161 &    144.7 &  $-$0.9 & $-$212.8 &     8.6 & 0.8 \\
52560.4184 &    151.3 &     5.4 & $-$225.9 &  $-$4.2 & 0.8 \\
52562.4949 & $-$184.9 &     3.4 &    219.0 &     8.6 & 1.2 \\
52564.3111 &    147.4 &     1.3 & $-$217.4 &     4.7 & 1.4 \\
52566.3018 & $-$189.0 &  $-$5.4 &    216.5 &    12.3 & 1.1 \\
52566.3165 & $-$183.5 &     1.1 &    216.5 &    10.9 & 1.0 \\
52568.3558 &    160.0 &     3.7 & $-$227.5 &     7.7 & 1.3 \\
52570.3545 & $-$184.4 &     5.4 &    216.5 &     4.4 & 1.3 \\
\hline \hline \end{tabular} \end{center} \end{table}

Radial velocities of the two stars were derived from the observed spectra using the two-dimensional cross-correlation algorithm {\sc todcor} (Zucker \& Mazeh 1994). In this method the observed spectra are cross-correlated against two template spectra simultaneously. Whilst in Paper\,I we used synthetic spectra from {\sc uclsyn} as templates, here we can use spectra obtained around the midpoint of secondary eclipse. As the eclipses are total, this contains only light from the primary star (and a negligible amount of contaminating light -- see section~\ref{ebop}). This template was used for both stars due to the similarity of their spectral characteristics, and allows us to avoid possible systematic errors due to the use of theoretical spectra. Best-fitting synthetic spectra were also generated and an independent solution was performed using {\sc todcor} to check our observed-template solution and determine the systemic velocities of the stars.

\begin{figure} \includegraphics[width=0.48\textwidth,angle=0]{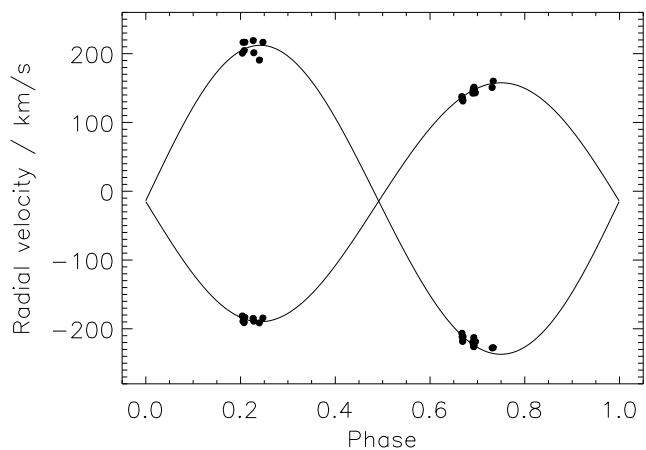} \\ \caption{\label{orbitplot} 
Spectroscopic orbit for V453\,Cyg from an {\sc sbop} fit to radial velocities from {\sc todcor}.} \end{figure}

{\sc todcor} was unreliable when the velocity separation of the stars was significantly lower than the combined rotational velocities of the stars. For this reason the best spectra were selected by eye, leaving sixteen covering the wavelength region of 4450--4715\,\AA\ and six spectra covering 4230--4500\,\AA. The hydrogen and helium lines at 4340\,\AA, 4471\,\AA\ and 4686\,\AA\ were masked to avoid significant errors due to the blending of broad spectral lines (Andersen 1975). The resulting radial velocities are given in Table~\ref{rvtable} and were fitted using {\sc sbop} in single-lined Lehman-Filh\'es mode. The orbital period, ephemeris time of reference, eccentricity and $\omega$ were fixed at values derived from the apsidal motion analysis. The results of this analysis are given in Table~\ref{spectable} and the final spectroscopic orbit is plotted in Fig.~\ref{orbitplot}.

\begin{table} \begin{center} \caption{\label{spectable} Parameters of the spectroscopic orbit derived fom V453\,Cyg using {\sc todcor} only on narrow lines. The systemic velocities were derived using {\sc todcor} and synthetic template spectra.}
\begin{tabular}{l r@{\,$\pm$\,}l r@{\,$\pm$\,}l} \hline \hline
                                          & \multicolumn{2}{c}{Primary} & \multicolumn{2}{c}{Secondary} \\ \hline
Orbital period $P$ (days)                 & \multicolumn{4}{c}{3.889825 (fixed)} \\
Reference time $T_0$ (HJD)                & \multicolumn{4}{c}{39340.6765 (fixed)} \\
Eccentricity $e$                          & \multicolumn{4}{c}{0.022 (fixed)} \\
Periastron longitude $\omega$ (\degr)     & \multicolumn{4}{c}{140.1 (fixed)} \\
Semiamplitude $K$ (\kms)                  & 173.7 & 0.8 & 224.6 & 2.0 \\
Systemic velocity (\kms)                  & $-$13.1 & 0.3 & $-$16.2 & 1.8 \\ \hline
Mass ratio $q$                            & \multicolumn{4}{c}{0.773 $\pm$ 0.008} \\
$a \sin i$ (\Rsun)                        & \multicolumn{4}{c}{30.59 $\pm$ 0.17} \\
$M \sin^3 i$ (\Msun)                      & 14.35 & 0.20 & 11.10 & 0.13 \\
\hline \hline \end{tabular}\end{center} \end{table}

Our velocity semiamplitudes are slightly larger, although in general consistent with, those found in the recent spectroscopic analyses of Popper \& Hill (1991), Simon \& Sturm (1994) and BMM97. This effect is probably because our radial velocities have been derived using only metal lines, whereas previous studies have relied mainly on helium lines. The effect of neglecting orbital eccentricity, however, is negligible, as can be seen from the two orbital solutions of the BMM97 radial velocities in Table~\ref{pubspecorbits}.


\section{Light curve analysis} \label{ebop}  

\begin{table*} \begin{center} 
\caption{\label{ldtable} Theoretical limb darkening coefficients (LDCs) for stars similar to the components of V453\,Cyg, compared to constraints from analysis of the light curves. A subscripted $A$ or $B$ denotes the LDC of the primary star or secondary star respectively. Filter designations are given in brackets.}
\begin{tabular}{l c r l c r l c r l} \hline \hline
Reference                          &\hspace{10pt}& $u_{\rm A} (U)$ & $u_{\rm B} (U)$ && $u_{\rm A} (B)$ 
                                                 & $u_{\rm B} (B)$ && $u_{\rm A} (V)$ & $u_{\rm B} (V)$ \\ \hline
Klinglesmith \& Sobieski (1970)                 && 0.340 & 0.321 &\hspace{10pt}& 0.317 & 0.302 && 0.268 & 0.249 \\ 
Wade \& Rucinski (1985)                         && 0.327 & 0.318 && 0.292 & 0.279 &\hspace{10pt}& 0.274 & 0.248 \\
van\,Hamme (1993)                               && 0.324 & 0.296 && 0.318 & 0.287 && 0.260 & 0.247 \\
D\'{\i}az-Cordov\'es, Claret \& Gim\'enez (1995)&& 0.374 & 0.357 && 0.376 & 0.357 && 0.334 & 0.316 \\
Claret (1998)                                   && 0.370 & 0.360 && 0.365 & 0.358 && 0.320 & 0.309 \\ 
Claret (2000)                                   && 0.423 & 0.384 && 0.420 & 0.381 && 0.374 & 0.334 \\ \hline
Largest LDCs which fit the light curves well    &&\multicolumn{2}{c}{0.5}&&\multicolumn{2}{c}{0.4}&&\multicolumn{2}{c}{0.35}\\
\hline \hline \end{tabular} \end{center} \end{table*}

\begin{table*} \begin{center} 
\caption{\label{lcfittable} Results of the light curve analysis for V453\,Cygni. The adopted values are the weighted means of the values determined from the individual light curves.}
\begin{tabular}{l l c r@{\,$\pm$\,}l r@{\,$\pm$\,}l r@{\,$\pm$\,}l c r@{\,$\pm$\,}l} \hline \hline
                                    &     &\hspace{10pt}& \multicolumn{2}{c}{$U$} & \multicolumn{2}{c}{$B$} &
                                            \multicolumn{2}{c}{$V$} &\hspace{10pt}& \multicolumn{2}{c}{Adopted} \\ \hline
Total number of datapoints          &     && \multicolumn{2}{c}{538} & \multicolumn{2}{c}{540} &
                                            \multicolumn{2}{c}{540} && \multicolumn{2}{c}{1618} \\             
Number used in solution             &     && \multicolumn{2}{c}{531} & \multicolumn{2}{c}{532} & 
                                            \multicolumn{2}{c}{534} && \multicolumn{2}{c}{1597} \\             
Linear limb darkening coefficient   & $u_{\rm A}$ && \multicolumn{2}{c}{0.324} & \multicolumn{2}{c}{0.318} & 
                                            \multicolumn{2}{c}{0.260} && \multicolumn{2}{c}{} \\
Linear limb darkening coefficient   & $u_{\rm B}$ && \multicolumn{2}{c}{0.296} & \multicolumn{2}{c}{0.287} & 
                                            \multicolumn{2}{c}{0.247} && \multicolumn{2}{c}{} \\ \hline
Primary radius ($a$)        & $r_{\rm A}$ && 0.2788 & 0.0021 & 0.2800 & 0.0014 & 0.2793 & 0.0014 && 0.2795 & 0.0009 \\
Secondary radius ($a$)      & $r_{\rm B}$ && 0.1781 & 0.0039 & 0.1785 & 0.0030 & 0.1811 & 0.0029 && 0.1794 & 0.0018 \\
Ratio of the radii                  & $k$ && 0.648  & 0.016  & 0.637  & 0.013  & 0.649  & 0.012  && 0.644  & 0.008  \\ 
Orbital inclination (degrees)       & $i$ && 89.9   & 1.3    & 88.2   & 1.2    & 89.0   & 1.1    && 89.0   & 0.7    \\
Surface brightness ratio            & $J$ && 0.938  & 0.014  & 0.953  & 0.010  & 0.948  & 0.015  && \multicolumn{2}{c}{} \\
Light ratio     & $L_{\rm B} / L_{\rm A}$ && 0.381  & 0.022  & 0.375  & 0.016  & 0.384  & 0.015  && \multicolumn{2}{c}{} \\
Third light (fraction of total light)&$L_3$&& 0.079 & 0.025  & 0.068  & 0.024  & 0.089  & 0.022  && \multicolumn{2}{c}{} \\ 
\hline \hline \end{tabular} \end{center} \end{table*}


\begin{figure*} \includegraphics[width=\textwidth,angle=0]{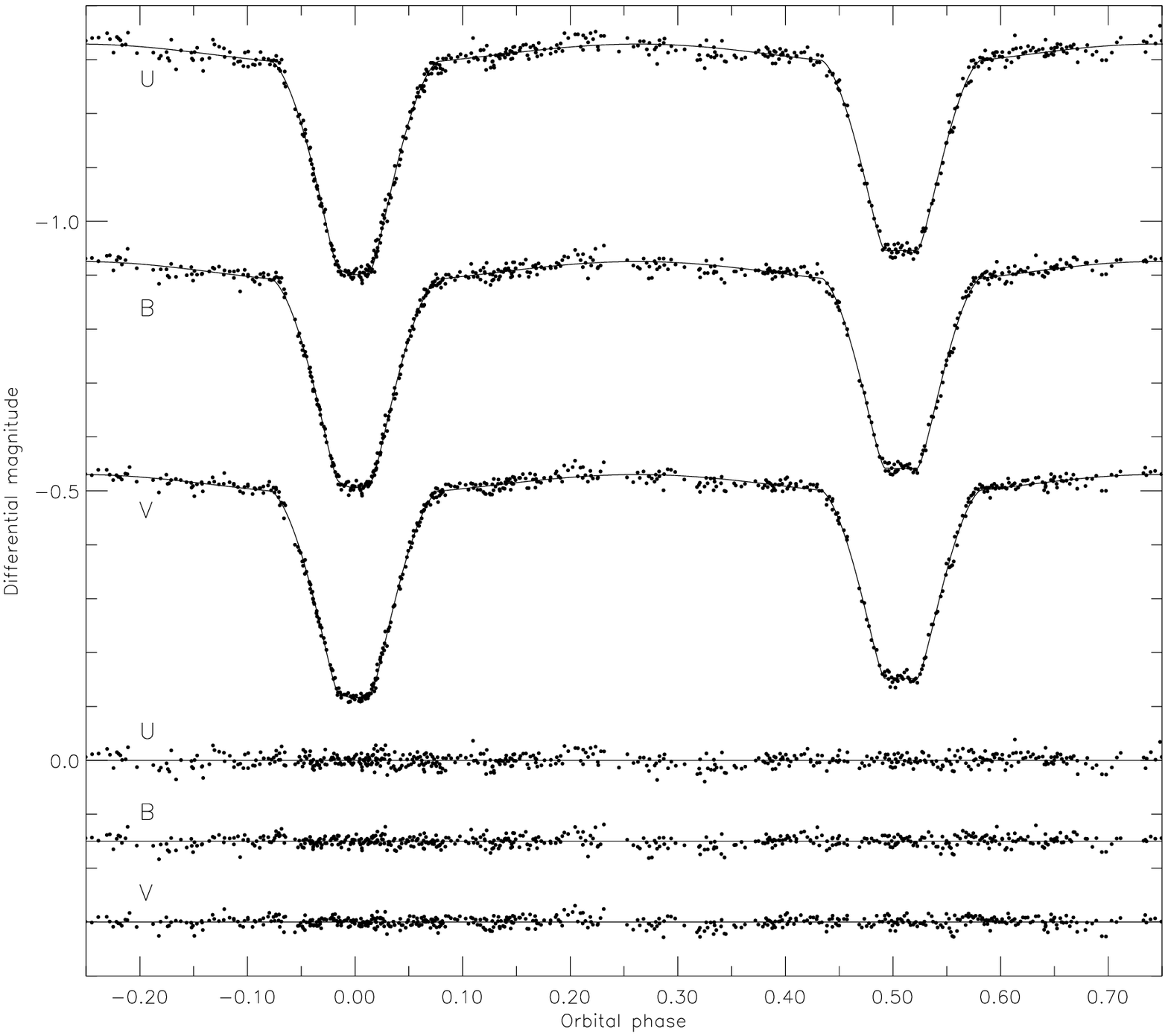} \\ \caption{\label{lcfitfigure} 
Observed phased light curves of V453\,Cyg with the best-fitting {\sc ebop} model light curves. The lower three curves show the residuals of the {\sc ebop} fits. For clarity the $B$ and $V$ residuals are offset by +0.15 and +0.3 magnitudes, respectively.} \end{figure*}

We have analysed the $UBV$ light curves taken from the work of Cohen (1974). As discussed in section~\ref{v453cyg}, these observations are not definitive, but for this totally eclipsing system they are able to provide accurate values of the individual stellar radii. We have used the simple and efficient NDE eclipsing binary model as implemented in the light curve analysis program {\sc ebop}\footnote{Eclipsing Binary Orbit Program written by Dr.\ Paul B.\ Etzel (\texttt{http://mintaka.sdsu.edu/faculty/etzel/}).} (Nelson \& Davis 1972, Popper \& Etzel 1981). In this model the stellar shapes are approximated by biaxial ellipsoids. The calculated oblatenesses of the best-fitting model for V453\,Cyg are within the limits of reliability for the {\sc ebop} code (Popper \& Etzel 1981). 

The light curves were phased with the sidereal period. Filter-specific linear limb darkening coefficients (LDCs) were taken from van Hamme (1993), gravity darkening exponents $\beta_1$ were fixed at 1.0 (Claret 1998) and the mass ratio was fixed at the spectroscopic value. After an initial solution was obtained, datapoints which showed residuals of more than 3$\sigma$ were rejected. These observations were clearly discrepant and their omission has not affected the derived parameter values but has lowered their uncertainties.

Initial investigation suggested that there is a small amount of third light, $L_3$, but acceptable solutions can be found without this effect. However, solutions with $L_3 \neq 0$ fit slightly better than solutions with $L_3 = 0$ for all three light curves, so third light has been included in all final solutions. If $L_3$ is neglected, a ratio of the radii lower by about 0.04 is required to reproduce the observed eclipse depths. The effect on the derived stellar radii is an increase in $R_{\rm A}$ by about 1\% and a decrease in $R_{\rm B}$ by about 4\%. These adjustments would bring our photometric solution into agreement with previous light curve analyses, which have all neglected third light and therefore may be systematically wrong.

Table~\ref{ldtable} shows several theoretical determinations of linear LDCs for stars having similar effective temperatures and suface gravities of the components of V453\,Cyg. We have evaluated the effect of a change in LDCs, on the parameters of the photometric solution. Table~\ref{ldtable} suggests that there is a variation of about 0.05 between different investigations of LDCs, so we perturbed the van Hamme (1993) values by this amount and refitted the light curves. The resulting errors have been added to the quoted uncertainties in Table~\ref{lcfittable} but are significant only for the surface brightness ratios. We have also determined the upper values of the LDCs for which light curve fits are not notably worse than our best fits, assuming that the limb darkening of both stars are adequately represented by the same LDC. This shows that the recent trend towards larger theoretical LDCs is producing values which are close to the observational maxima for our $B$ and $V$ light curve solutions.

The best-fitting light curves are compared to the observations in Fig.~\ref{lcfitfigure}. The residuals of the fit are also shown, and some minor systematic trends are noticeable. Whilst the {\sc ebop} light curve model is adequate to fit the current photometric data, definitive light curves may require a more sophisticated treatment such as that contained in the Wilson-Devinney code (Wilson \& Devinney 1971, Wilson 1993), which has a better representation of limb darkening and the reflection effect. V453\,Cyg is a good system for the determination of observational LDCs due to the long totality of its primary eclipse. The Cohen (1974) light curves are not of sufficient quality to determine LDCs here; definitive light curves will be required.

\subsection{Error analysis}

We have modified the {\sc ebop} code to use the Levenberg-Marquardt minimisation algorithm ({\sc mrqmin}; Press \etal\ 1992) to find the best fit to the data. Whilst {\sc mrqmin} allows calculation of the formal errors of the adjusted light curve parameters, it is known that these uncertainties are very optimistic when some parameters are significantly correlated. Correlations are generally small for systems which exhibit total eclipses, except for systems with third light. Orbital inclination and $L_3$ can be strongly anticorrelated as both have a significant dependence on the depth of the eclipses. Robust estimation of uncertainties must include an assessment of parameter correlations for the physical characteristics of the system under investigation.

We have used the method of bootstrapping to evaluate the uncertainties and correlations of the parameters derived from the light curve analysis. After the best fit was determined for each light curve, a synthetic light curve was evaluated at the phases of observation of the real light curve. We added observational noise of the same magnitude as the real light curve and refitted the synthetic light curve with {\sc ebop}. This process was undertaken ten thousand times for each observed light curve.

\begin{figure} \includegraphics[width=0.48\textwidth,angle=0]{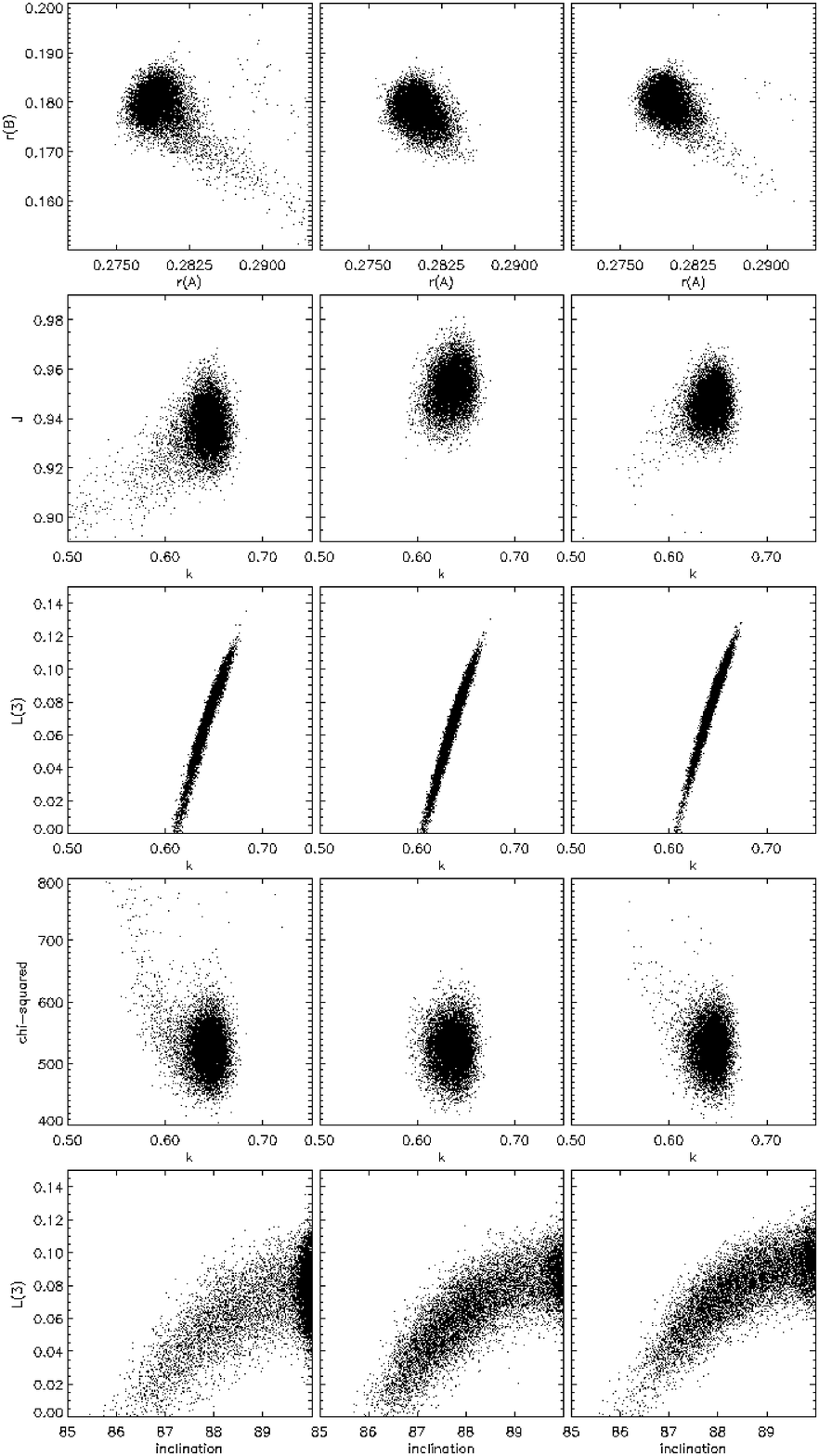} \\ \caption{\label{lcerrplot} Sample distributions of the best-fitting parameters evaluated during the bootstrapping analysis. The units and parameter symbols are as in Table~\ref{lcfittable}. Each distribution between two parameters is shown for the $U$ (left), $B$ (middle) and $V$ (right) light curves.} \end{figure}

It is important to understand what information these bootstrapping simulations actually provide. Once a best fit is found, the distributions of the ten thousand evaluations of various parameters give us the parameter uncertainties and their correlations, based on the best fit, the phases of observation, and the observational scatter of the real light curves. This is a valid method of analysis if the best fits are close to the true characteristics of the dEB. The reality of this assumption can be assessed using independent solutions of different light curves, for example the $U$, $B$ and $V$ observations here. The bootstrapping analysis then serves as an indication of the validity of uncertainties estimated from the interagreement of different light curves.

Sample plots of the distributions of different parameter values are shown in Fig.~\ref{lcerrplot}. It is notable that the ratio of the radii, $k$, and the ratio of the surface brightnesses, $J$, are not correlated due to this system exhibiting total eclipses. However, $k$ and $L_3$ show a very strong correlation and illustrate why $L_3$ has not been included in previous light curve analyses. This effect is because, for a given value of $k$, a well-defined value of $L_3$ is required to fit the well-determined eclipse depths. 


The best-fitting photometric parameters of V453\,Cyg, their 1$\sigma$ uncertainties, and the final adopted parameters are given in Table~\ref{lcfittable}. The adopted parameters were determined using weighted means of the values determined from the individual light curves, and their uncertainties are similar to but larger than the standard deviations of the individual values.


\section{Absolute dimensions and comparison with stellar models}      \label{dimensionsmodels}

\begin{table} \begin{center} \caption{\label{absolutedimensions}
Absolute dimensions of the detached eclipsing binary V453\,Cygni in the open cluster NGC\,6871. 
\newline $^*$\,Calculated using the combined magnitude and flux ratio in the $V$ filter, the assumed cluster distance modulus and reddening, and the canonical reddening law $A_V = 3.1\EBV$.
\newline $^\dag$\,Calculated using the effective temperature--bolometric correction calibration of Flower (1996)
\newline $^\ddag$\,Taken from Olson (1984).
\newline \Veq\ and \Vsync\ refer to the equatorial and the synchronous rotational velocities respectively.}
\begin{tabular}{l r@{\,$\pm$\,}l r@{\,$\pm$\,}l} \hline \hline
                                    & \multicolumn{2}{c}{V453\,Cyg A} & \multicolumn{2}{c}{V453\,Cyg B} \\ \hline
Cluster age $\log\tau$ (years)      & \multicolumn{4}{c}{6.3 to 6.7}                \\ 
Cluster distance modulus            & \multicolumn{4}{c}{11.65 $\pm$ 0.07}          \\ \hline
Orbital period (days)               & \multicolumn{4}{c}{3.889825 $\pm$ 0.000017}   \\
Mass ratio $q$                      & \multicolumn{4}{c}{0.773 $\pm$ 0.008}         \\
Mass (\Msun)                        & 14.36     & 0.20      & 11.11     & 0.13      \\
Radius (\Rsun)                      & 8.551     & 0.055     & 5.489     & 0.063     \\
\logg\ (\cms)                       & 3.731     & 0.012     & 4.005     & 0.015     \\ 
Effective temperature (K)           & 26\,600   & 500       & 25\,500   & 800       \\
$M_V$$^*$ (mag)                     & $-$4.44   & 0.38      & $-$3.39   & 0.39      \\
Luminosity$^\dag$ ($\log L/L_\odot$)& 4.69      & 0.21      & 4.24      & 0.28      \\ \hline
\Veq$^\ddag$ (\kms)                 & 107       & 9         & 97        & 20        \\
\Vsync\ (\kms)                      & 111.3     & 0.7       & 71.4      & 0.8       \\
Systemic velocity (\kms)            & $-$13.1   & 0.3       & $-$16.2   & 1.8       \\ \hline
Apsidal motion period (yr)          & \multicolumn{4}{c}{66.4 $\pm$ 1.8}            \\
$\log k_2$                          & \multicolumn{4}{c}{$-$2.226 $\pm$ 0.024}      \\
\hline \hline \end{tabular} \end{center} \end{table}

The derived physical parameters for the component stars of V453\,Cyg have been collected in Table~\ref{absolutedimensions}; the masses and radii of the two stars have been measured to accuracies better than 1.4\%. The radius of the primary star is extremely well determined because it depends mainly on the duration of totality during secondary eclipse. The rotational velocities of the stars are slightly uncertain: using our spectroscopic data we have been unable to derive values more accurate than those of Olson (1984). The primary star rotates synchronously with the orbital velocity but the secondary rotates somewhat faster (although with a large uncertainty). 

The timescale for orbital circularisation is about 1.5\,Gyr (Zahn 1977; Hilditch 2001) which is consistent with the presence of a small but well-defined eccentricity. The timescale for synchronisation is approximately 3.0\,Myr (Zahn 1977; Hilditch 2001) and the rotational velocity of the secondary star suggests that V453\,Cyg is younger than this. As we have measured the age of V453\,Cyg to be $10.0 \pm 0.2$\,Myr by comparison with the predictions from theoretical models, we expect that this slight discrepancy is due to the secondary rotational velocity being quite uncertain.

\subsection{Stellar model fits}

The absolute parameters of the components of V453\,Cyg have been compared to the predictions of stellar models from four different groups:-- (1) Claret (1995), hereafter referred to as the Granada95 models, (2) Bressan \etal\ (1993), hereafter referred to as the Padova93 models (the more recent Padova models of Girardi \etal\ (2000) only extend to stellar masses of 7\Msun), (3) Schaller \etal\ (1992), hereafter referred to as the Geneva92 models, and (4) Pols \etal\ (1998) with modifications\footnote{For details of these modifications please refer to the website of O.\ R.\ Pols: \texttt{http://www.astro.uu.nl/$\sim$pols/}}, hereafter referred to as the Cambridge2000 models. For each set of models we have interpolated over age using cubic spline functions and plotted the resulting predictions in the mass--radius and logarithmic effective temperature--surface gravity diagrams. Comparisons with the properties of V453\,Cyg were performed simultaneously in both diagrams and the two stars were assumed to have the same age and chemical composition (as expected for close binary stars). The Granada95, Padova93 and Geneva92 models all include a moderate amount of convective core overshooting (although with different formalisations). Happily, the Cambridge2000 models are available both with and without a moderate amount of overshooting, allowing us to test whether the inclusion of this effect provides a better fit to the observational data.

\begin{figure*} \includegraphics[width=\textwidth,angle=0]{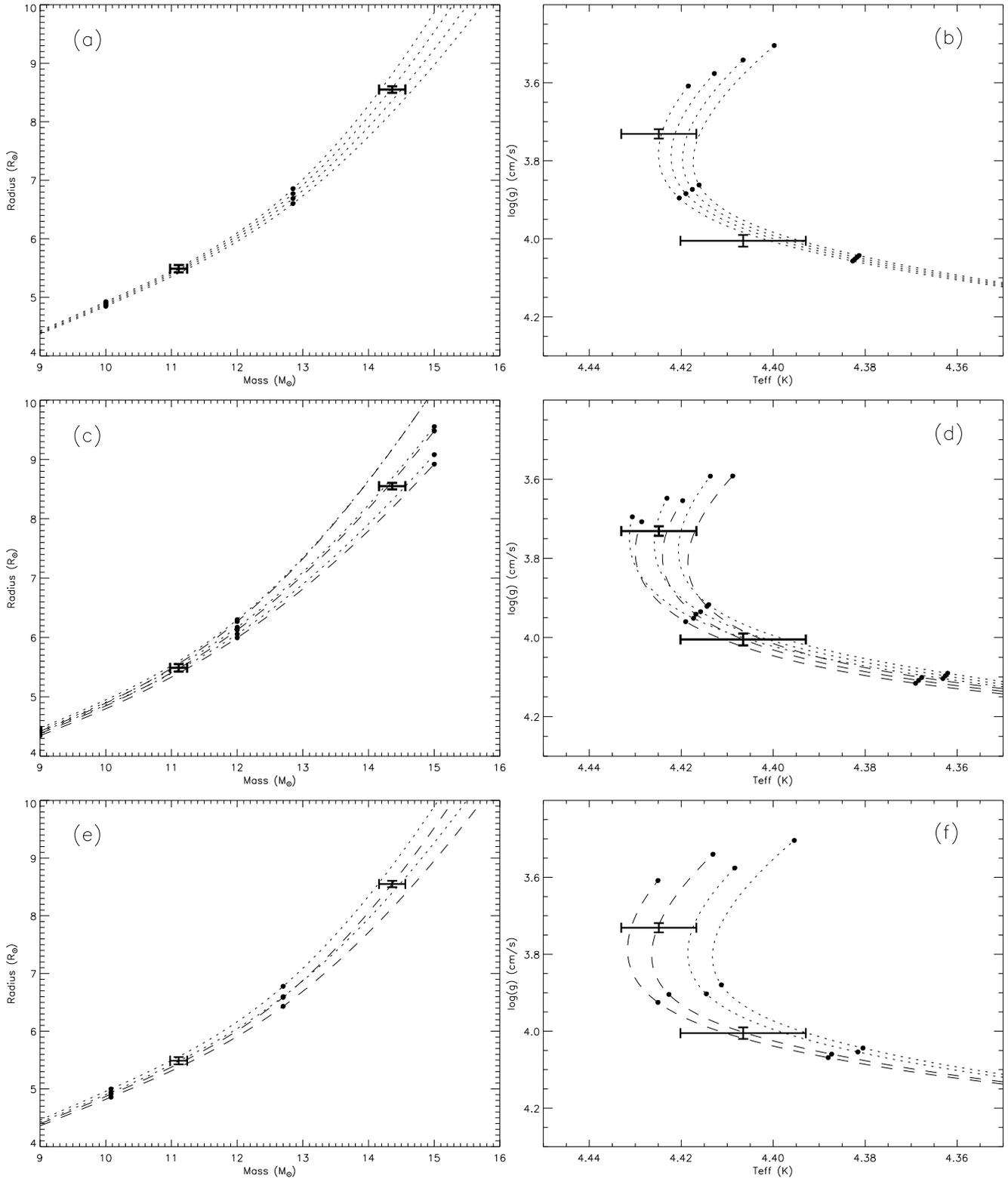} \\ \caption{\label{modelfit} Comparison between stellar models and the absolute dimensions of V453\,Cyg in the mass--radius and the \Teff--\logg\ diagrams. Isochrones have been plotted to represent the stellar models, with circles showing their points of evaluation. Broken lines have been plotted by interpolating over mass using cubic splines. Panels (a) and (b) show the Granada stellar models for ($X$,$Y$) = (0.02,0.28) and ages of 9.7, 9.9, 10.1 and 10.3 Myr (radii increase and \Teff s decrease as age increases). Panels (c) and (d) show the Padova models for ($X$,$Y$) = (0.02,0.28) (dotted lines) and Geneva models for ($X$,$Y$) = (0.02,0.30) (dashed lines) for ages of 9.4, 9.8 and 10.2 Myr. Panels (e) and (f) show the Cambridge models for ($X$,$Y$) = (0.02,0.28) with overshooting (dashed lines) and without overshooting (dotted lines, for ages of 9.8 and 10.2 Myr (with overshooting) or 9.4 and 9.8 Myr (with no overshooting).}  \end{figure*}

Panels (a) and (b) of Fig.~\ref{modelfit} show the parameters of V453\,Cyg compared to the predictions of the Granada95 models. A good fit is obtained for an age of 9.9\,Myr and a chemical composition of ($Z$,$Y$) = (0.02,0.28) (i.e., normal helium abundance). Attempts to fit the stars with a higher or lower helium abundance (Claret 1995) or metal abundances of $Z = 0.01$ (Claret \& Gim\'enez 1995) or $Z = 0.03$ (Claret 1997) were unsuccessful.

The predictions of the Padova93 and the Geneva92 models are compared to the parameters of V453\,Cyg in panels (c) and (d) of Fig.~\ref{modelfit}. The two sets of model predictions are plotted for $Z = 0.02$ and the same three ages, so are directly comparable apart from a slight difference in the assumed helium abundance ($Y = 0.28$ for the Padova93 models and $Y = 0.30$ for the Geneva92 models). It is notable that the two sets of models agree very well not only with each other but with the comparable Granada95 models discussed above. Whilst both the Padova93 and Geneva92 models fit the components of V453\,Cyg best for an age of 9.8\,Myr, a marginally better fit is provided by the slightly higher predicted \Teff\ values of the Padova93 models. Attempts were also made to fit the components of V453\,Cyg using metal abundances higher [$Z = 0.05$ for the Padova93 models (Fagotto \etal\ 1994a); $Z = 0.04$ for the Geneva92 models (Schaerer \etal\ 1993b)] and lower [$Z = 0.008$ for the Padova93 models (Fagotto \etal\ 1994b); $Z = 0.008$ for the Geneva92 models (Schaerer \etal\ 1993a)] without any success.

The Cambridge2000 model set differs from the other model sets considered here in that it is available with and without a moderate amount of convective core overshooting, but does not include any mass loss; this should be unimportant for these stars. Panels (e) and (f) of Fig.~\ref{modelfit} show the parameters of V453\,Cyg compared to the predictions of the Cambridge2000 models. Both overshooting and standard-mixing isochrones are plotted for an age of 9.8\,Myr for comparison. The overshooting models are also plotted for the best-fitting age of 10.2\,Myr and the standard-mixing models are plotted for their best-fitting age of 9.4\,Myr. The overshooting models are notably more successful than the standard-mixing models, which predict \Teff\ values which are slightly too low for the stars of V453\,Cyg. As with the Padova93 and Geneva92 models, we were unable to perform fits for different helium abundances as such models have not been published.

The above comparisons demonstrate that a good agreement has been reached between different sets of theoretical evolutionary models for stars similar to the components of V453\,Cyg. All sets of models were successful in fitting the observations for an age of $10.0 \pm 0.2$\,Myr and solar metal and helium abundances. We also attempted to fit models to the absolute dimensions of V453\,Cyg derived with zero third light. This changes the radii to 8.649 and 5.250\Rsun\ and the surface gravities to $\logg = 3.723$ and 4.045 (\cms), with other quantities, and the uncertainties, unaffected. Using the Granada95 models we were able to achieve a fit in the mass--radius plane for low metal abundance ($Z = 0.01$), high helium abundance ($Y = 0.36$) and an age of 8.2\,Myr. However, the \Teff\ values were predicted to be 2000\,K greater than observed, and a combination of low metal abundance and high helium abundance does not agree with the predictions of Galactic chemical evolution theory (see e.g., Binney \& Merryfield 1998). We were unable to find a simultaneous fit in the mass--radius and \Teff--\logg\ diagrams for the Geneva92, Padova93 or Cambridge2000 models.

BMM97 successfully fitted the Geneva92 theoretical models to the observed masses and luminosities of the components of V453\,Cyg. As they did not compare stellar radii, they were not subject to errors from the assumption of no third light. The luminosities of V453\,Cyg are also much more uncertain than the radii derived here, so fitting in the $\log\Teff - M_{\rm bol}$ plane allows a wider range of predictions to fit the observed data.

\subsection{Comparison between the observed apsidal motion constant and theoretical predictions}

The internal structure constant $\log k_2$ is a measure of the density concentration of components of a dEB which shows apsidal motion. As discussed in Claret \& Gim\'enez (1993), the observed density concentration coefficient can be calculated from the apsidal motion period using the equation \[ \overline{k_2^{\rm ~obs}} = \frac{1}{c_{21} + c_{22}} \frac{P}{U} \]
where $P$ and $U$ are the orbital period and apsidal period and the constants $c_{2i}$ are weights which depend on the stellar mass, radius and rotational velocity ($i$=1 refers to the primary star and $i$=2 refers to the secondary). The individual theoretical density concentration coefficients, $k_{2i}$, must be combined using the equation 
\[ \overline{k_2^{\rm ~theo}} = \frac{c_{21} k_{21} + c_{22} k_{22}}{c_{21} + c_{22}} \]
to find the weighted average coefficient which is directly comparable the observed value. 

We must remember that the observed $\log k_2$ contains not only the Newtonian contribution to apsidal motion, which is due to the stars being distorted extended shapes, but a small amount of general relativistic apsidal motion. This last contribution can be straightforwardly calculated using 
\[ \dot\omega = 5.45 \times 10^{-4} \frac{1}{1-e^2} \left(\frac{M_1 + M_2}{P}\right)^{2/3} \]
(Gim\'enez 1985) where $\dot\omega$ is the apsidal motion rate in degrees per cycle, $e$ is the orbital eccentricity, $P$ is the anomalistic period (days) and $M_i$ are the stellar masses (\Msun).

The tabulated results for the Granada95 theoretical models include predicted internal structure constants which can be compared with our observations. The general relativistic contribution to the apsidal motion of V453\,Cyg is approximately 6\% of the total apsidal motion rate, and its subtraction gives a Newtonian density concentration coefficient of 
\[ \log \overline{k_2^{\rm ~obs}} = -2.254 \pm 0.024 \]
(neglecting any uncertainty in calculation of the general relativistic contribution). The theoretical weighted average coefficient was calculated using individual $k_{2i}$ values from interpolation, in the table of model predictions, to the observed masses and best-fitting age of V453\,Cyg. This value is dominated by the primary star, as it is more distorted due to its significantly larger radius, and is 
\[ \log \overline{k_2^{\rm ~theo}} = -2.255 \]
The agreement with observations is excellent. This agreement is particularly important for assessing the assumed amount of convective core overshooting, on which theoretical values of $\log k_2$ have a significant dependence (Claret \& Gim\'enez 1991).


\section{Membership of the open cluster NGC\,6871}

V453\,Cyg is traditionally considered to be a member of the NGC\,6871 open cluster, and appears on the cluster main sequence in all photometric diagrams. Further proof of membership comes from its systemic velocity, $-13.2 \pm 0.3$\kms\ using a weighted mean of systemic velocities calculated for each star. This agrees well with the value of $-15 \pm 6$\kms\ quoted by Hron (1987) and the radial velocity of the cluster member NGC\,6871\,13 which was measured to be $-14.6 \pm 2.7$\kms\ using the same instrumental setup as we used for V453\,Cyg. However, the radial velocity of NGC\,6871 is given as $-7.7 \pm 3.2$\kms\ by Rastorguev \etal\ (1999), which differs from our value for V453\,Cyg by 1.6$\sigma$. The proper motion of V453\,Cyg is consistent with cluster membership (Perryman \etal 1997).

Massey \etal\ (1995) give an age of 2 to 5\,Myr for the members of NGC\,6871 with the earliest spectral types, but their photometric diagrams contain somewhat evolved 15\Msun\ stars which are also claimed to be cluster members. This suggests that the stars in NGC\,6871 have either a spread in ages or were created by two distinct bursts of star formation. We cannot currently distinguish between the two possibilities; the age of $10.0 \pm 0.2$\,Myr derived for V453\,Cyg using theoretical models is consistent with both evolutionary scenarios.


\section{Discussion}    \label{discussion}

We have derived the absolute dimensions of the components of the high-mass detached eclipsing binary V453\,Cygni, a member of the open cluster NGC\,6871. 

Effective temperatures were found using the helium ionisation balance derived from high-resolution spectra, which also suggest an enhanced photospheric helium abundance relative to solar. Radial velocities were derived from the spectra using only the weak spectral lines and the {\sc todcor} two-dimensional cross-correlation algorithm. 

The apsidal motion rate of the system has been determined using an extended version of the photometric method of Lacy (1992), which includes times of minimum light and spectroscopic determinations of eccentricity and $\omega$. The apsidal period is well constrained, and allow the derivation of eccentricity and $\omega$ to a greater accuracy than possible with the light curves and radial velocity curves. 

We have reanalysed the $UBV$ light curves of Cohen (1974) in order to determine the radii of the components of the dEB. The best-fitting parameters include a small amount of third light, which was previously undetected. Robust parameter uncertainties were derived using bootstrapping, allowing us to quantify and illustrate the effect of correlations between different photometric parameters. The ratio of the radii and the amount of third light are strongly correlated, due to their dependence on the depths of the eclipses; previous photometric studies which did not include third light are systematically biased towards values of the stellar radii which are 1\% higher and 5\% lower for primary and secondary respectively.

The accurate absolute dimensions presented here allow V453\,Cyg to be added to the list of dEBs with the best-determined values of mass, radius and effective temperature (Andersen 1991). However, our analysis would clearly be much improved with better observational data. The inclusion of only a few new times of minima would greatly increase the accuracy of the results of the apsidal motion analysis, and more accurate rotational velocities would allow a more accurate derivation of the internal structure constant $\log k_2$. A definitive spectroscopic orbit will require observations with a higher signal-to-noise than those presented here, and should give masses determined to accuracies of better than 1\%. Definitive light curves of the system would allow determination of the limb darkening coefficients for both stars, providing an important test of model atmosphere codes.

The absolute masses, radii and effective temperatures of the components of V453\,Cyg have been compared to several stellar models in the mass--radius and $\log\Teff$--\logg\ planes, assuming the same age for both stars. Not only is there impressive agreement betwen different theoretical models, all model sets are able to fit the observational data for a solar helium and metal abundance. Moreover, the Granada models (Claret 1995) provide a perfect match to the observed apsidal motion rate once the relativistic contribution has been subtracted from the overall effect. Stellar models have for a long time appeared to predict that the central condensations of stars are lower than that found using observations of apsidal motion (e.g., Clausen, Gim\'enez \& Scarfe 1986, Barembaum \& Etzel 1995). This apparent discrepancy has been reduced by the discovery that the internal structure constants change significantly through a star's evolution. The current generation of theoretical models, incorporating OPAL opacity data (Rogers \& Iglesias 1992), are in good agreement with observations. It is noticeable that some observers have not removed the general relativistic effect from their observed $\log k_2$ values before comparison with theory; in many cases this will have a negligible effect but for the stars of V453\,Cyg it causes about 6\% of the observed apsidal motion, changing $\log k_2$ by an amount similar to its uncertainty.

The normal helium abundance implied by stellar model fits also conflicts with the slight overabundance noted in our spectral synthesis analysis. We note that the photospheric helium abundance is not directly comparable to the initial internal helium abundance used in model calculations. 

Fits to the Cambridge stellar models support the inclusion of a moderate amount of overshooting in most stellar evolutionary models. Whilst models without overshooting were able to fit the masses and radii of the stars, the predicted effective temperatures are slightly lower than that determined from the helium ionisation balance.

The stellar models were extremely successful in fitting the absolute dimensions and effective temperatures of a high-mass slightly-evolved dEB, with component masses and radii differing by ten and twenty-five times their combined uncertainties, respectively. For observational stellar astrophysicists, this fact implies that we must either observe systems so thoroughly that their masses and radii are known to accuracies of 0.5\% and the effective temperatures to 2\%, or target particular types of stars to critique the success of one set of stellar models compared to another. Such targets include low-mass, high-mass, pulsating, and Population II stars, as well as eclipsing systems found in Local Group galaxies. Eclipsing binaries in open clusters can satisfy this requirement if the cluster they belong to is well-studied or otherwise interesting (Paper\,I), and further observations are being undertaken towards this goal.


\section*{Acknowledgements}

The authors would like to thank Ansgar Reiners, Roger Diethelm and Onno Pols for their assistance in parts of this analysis. We are particularly grateful to David Holmgren for making his apsidal motion code available, and to the referee for a very prompt and useful reply. We would also like to thank Paul Etzel and Imre Barna B\'{\i}r\'o for helpful discussions, and Liza van Zyl for critical reading of the manuscript.

JS acknowledges financial support from PPARC in the form of a postgraduate studentship. The authors acknowledge the data analysis facilities provided by the Starlink Project which is run by CCLRC on behalf of PPARC. The following internet-based resources were used in research for this paper: the WEBDA open cluster database; the ESO Digitized Sky Survey; the NASA Astrophysics Data System; the SIMBAD database operated at the CDS, Strasbourg, France; and the VizieR service operated at the CDS, Strasbourg, France.


\end{document}